\documentstyle[prl,multicol,aps,epsf]{revtex}
\begin{document}

\draft

\title{Coulomb Drag as a Signature
of the Paired Quantum Hall State}

\author{Fei Zhou$^{a}$ and  Yong Baek Kim$^{b,c}$}

\address{$^{a}$Physics Department, Princeton University, Princeton, NJ 08544}
\address{$^{b}$Department of Physics, The Pennsylvania State University,
University Park, PA 16802}
\address{$^{c}$Institute for Theoretical Physics, 
University of California, Santa Barbara, CA 93106}

\maketitle

\begin{abstract}
Motivated by the recent Coulomb drag experiment
of M. P. Lilly, et.al, Phys. Rev. Lett. {\bf 80},
1714 (1998),
we study the Coulomb drag in a  
two-layer system with Landau level
filling factor $\nu = 1/2$.
We find that the drag conductivity 
in the incompressible paired quantum Hall state 
at zero temperature can be finite. The drag conductivity is also  
greatly enhanced above $T_c$,
at which the transition between 
the weakly-coupled compressible liquids 
and the paired quantum Hall liquid takes place.
We discuss the implications
of our results for the recent experiment.
\end{abstract}

\begin{multicols}{2}

A double-layer system of two dimensional electron gases (2DEG)
allows a novel measurement of scattering 
mechanism\cite{eisenstein,gramila}.
If there is no tunneling between the two layers, 
momentum can be transferred 
only via electron-electron scattering 
due to the interlayer Coulomb interaction. 
As a result, if a current is driven through one of the subsystems 
(active layer), then another current is induced in the other 
system (passive layer).
The magnitude of the induced current is a measure
of the interlayer scattering rate.
In real experiments, an induced voltage is measured in the passive 
layer where no current flows.
The ratio between the measured voltage in the passive layer
and the driven current in the active layer is the 
so-called transresistivity or the drag resistivity.
In the case of a double-layer 2DEG system in the absence of
external magnetic field, 
only the quasiparticles
within an energy band of width $kT$ near the Fermi surface 
participate in scattering processes. This
leads to a $T^2$
temperature dependence of the drag resistivity at 
low temperatures\cite{macdonald,kamenev}. 
When the filling fraction
becomes one-half 
in the presence of high magnetic fields, 
the 2DEG in each layer
supports a novel form of compressible liquid\cite{HLR}.
Namely, the quasiparticles of the  
half-filled Landau level are composite fermions,
which are the electrons with Chern-Simons flux quanta attached to
them. Chern-Simons field fluctuations 
due to the density fluctuations of electrons lead to a more singular 
low energy interlayer scattering rate\cite{kim,stern,sahki}.
Theoretically it was found that the drag resistivity goes
as $T^{4/3}$ for a pure system and $T^{2}\ln T$ for a diffusive system.

Recently, Coulomb drag measurement was done for double layers
of half-filled Landau levels\cite{eisenstein}. 
In the experiment, it was indeed found that the drag resistivity is
much enhanced compared to that of 2DEG in the zero magnetic field. 
However, even though the temperature dependence can be fit to $T^{4/3}$
for a range of intermediate temperatures, the experiment revealed
much richer physics at low temperatures. 
It was observed that a) the drag resistivity has a minimum
at a certain temperature below which the drag becomes very
sensitive to disorder and the applied current;
b) the drag resistivity {\em seems} to be finite at the zero temperature.

Motivated by this experiment,  
we study the Coulomb drag in the paired quantum Hall state
limit. 
Incompressible paired quantum Hall states
with two electron species were suggested some years ago, based on 
both numerical simulations and effective action approaches
\cite{Halperin,Haldane,Wen,bonesteel}. 
In particular, it was suggested that, in double layers of 
Landau level filling factor $\nu=1/2$, composite fermions 
in one layer can establish the pairing 
correlation with composite fermions in the other layer 
below a certain temperature, $T_c$\cite{bonesteel}.
Though such a pairing correlation of 
composite fermions, which is responsible
for the incompressibility of the paired quantum Hall state,
{\it does not} lead to conventional 
long-rang order of electrons,  
it does introduce short-range pairing correlation 
of electrons, i.e. quantum fluctuations of electron pairs.
The question arises: {\em How does the short-range pairing correlation 
developed by electrons in the incompressible phase 
affect the Coulomb drag?}

In this letter, we study the transport properties of
this incompressible phase and the temperature dependence of
various transport coefficients. 
We find the following results. 
A) At $T=0$, the drag conductivity can be finite 
in the incompressible paired quantum Hall state. 
Its temperature dependence for $T < T_c$ 
strongly depends on disorder.
B) Above $T_c$, the drag conductivity
is enhanced by $ \sigma^{xx}_{12} \propto (e^2 /\hbar) 
(k_F l)^{-2} T/(T-T_c)$.
Here $k^{-1}_F=l_B$ is the Fermi wavelength and
much shorter than the mean free path of the electrons $l$. 
The Hall drag conductivity exhibits a similar enhancement 
near $T_c$. We also obtain the drag resistivities below and
above $T_c$.
We discuss the implications of these results to 
the experiment and suggest that the observed anomaly 
could be interpreted as a signature 
of the formation of an incompressible double-layer paired  
quantum Hall state at low temperatures. 

In the framework of composite fermion theory\cite{HLR},   
the response functions of electrons 
can be expressed in terms of those of composite fermions;
as a consequence, the in-plane conductivity, Hall conductivity,
and drag conductivity, Hall drag conductivity
can be expressed in terms of the composite fermion 
polarizabilities,

\begin{eqnarray}
&&\sigma^{xx}_{11, 12}=\frac{Im}{2\Omega}
\lim_{\Omega,{\bf Q}\rightarrow 0}
\left(  
\frac{\pi^{cc}_{-}}{1+c^2 \pi^{cc}_-
\pi^{dd}_-}\pm
\frac{\pi^{cc}_{+}}
{1+c^2 \pi^{cc}_+\pi^{dd}_+}
\right) \nonumber \\
&&\sigma^{xy}_{11, 12}=\frac{e^2}{8\pi\hbar}
\lim_{\Omega, {\bf Q}\rightarrow 0}
\left( \frac{c^2\pi^{dd}_{-} \pi^{cc}_{-}}
{1+c^2 \pi^{cc}_-\pi^{dd}_-}\pm
\frac{ c^2 \pi^{dd}_{+}\pi^{cc}_{+}}
{1+c^2 \pi^{cc}_+\pi^{dd}_+}
 \right).  
\end{eqnarray}
where 
${\pi^{cc}_{\pm}}
={\pi^{cc}_{11}(i\Omega, {\bf Q})} \pm{\pi^{cc}_{12}(i\Omega, {\bf Q})}$,
${\pi^{dd}_{\pm}}
={\pi^{dd}_{11}(i\Omega, {\bf Q})} \pm{\pi^{dd}_{12}(i\Omega, {\bf Q})}$.
$\pi_{\alpha\beta}^{dd}$ and 
$\pi_{\alpha\beta}^{cc}$ denote the density-density and
current-current polarization matrices of composite fermions respectively,
defined in the space of the layer index $\alpha,\beta=1,2$.
$c=i4\pi/Q$ comes from the Chern-Simons transformation.
In the incompressible double-layer
quantum Hall liquid limit,  we introduce the Green functions of 
composite fermions defined in a generalized Nambu space
in Matsubara representation\cite{bonesteel} 
\begin{eqnarray}
&&\hat{G}=\left(
\begin{array}{cc}
\tilde{G},  \tilde{F} \\
\tilde{F}^{+},   -\tilde{G}
\end{array}
\right),\ 
\tilde{G}=\left (
\begin{array}{cc}
{G_{11}},  0 \\
0,  {G_{22}}
\end{array}
\right),\ 
\tilde{F}=\left(
\begin{array}{cc}
0, {F_{12}}\\
{F_{21}},  0 
\end{array}
\right).
\nonumber \\
\end{eqnarray} 
Here $\tilde{G}$ and $\tilde{F}$ are defined in layer-index space, 

\begin{equation}
G_{11, 22}=
\frac{i\omega +\xi_{\bf p}}{\omega^2 + \Delta^2 +\xi_{\bf p}^2},
F_{12, 21}=\frac{\Delta}{\omega^2 + \Delta^2 +\xi_{\bf p}^2},
\end{equation}
where $\xi_{\bf p}={\bf p}^2/2m -\epsilon_F$ and $\omega=(2n+1)\pi T$. 
We first consider the clean limit $\tau \Delta \gg 1$, where $\tau$ is the
elastic mean free time.
${\Delta}$ is determined by the self consistent equation 
${\Delta(\omega)}=T\sum_{\Omega} g(\Omega) F_{12}(\omega -\Omega)$,
where $g(\Omega)$ is the interaction constant 
in the interlayer particle-particle channel.  
The external field vertices are renormalized
accordingly\cite{AGD}, as shown in Fig.1a.
To simplify the calculation we neglect the energy dependence
of $g_{12}(\Omega)$ and $\Delta$. We also ignore 
intralayer Fermi liquid renormalization effects.
To the leading order in the small parameter
$\Delta/\epsilon_F$, when $v_F Q, \Omega \ll \Delta$, 
the diagrams in Fig.1a yield 
\begin{eqnarray}
\hat{\Gamma}_0=\hat \tau_3 + {\hat \tau_2} \frac{\Omega \Delta}{\Omega^2 + 
{v_s}^2{Q}^2}, \ \  
& \hat{\bf \Gamma}={\bf v}_F. 
\end{eqnarray}
where $v_s=v_F\alpha_0(T)/\sqrt{2}$,
$v_F$ is the Fermi velocity, and $\alpha_0(T)$ is a temperature
dependent constant. 
$\alpha_0(T=0)=1$ and for $T \sim T_c$ (where $\Delta \sim 0$),
$\alpha_0(T)=\sqrt{7\zeta(3)\Delta/2\pi^2 T} \ll 1$.
$\zeta(x)$ is the Riemann Zeta function.
$\hat{\tau}_2=\left(
\begin{array}{cc}
0, & \tau_1 \\
-\tau_1, & 0
\end{array}
\right)$, 
$\hat{\tau}_3=\left(
\begin{array}{cc}
\tau_0, & 0 \\
0, & -\tau_0
\end{array}
\right)$, 
where $\tau_0$, $\tau_1$ are the unity matrix and x-component
Pauli matrix in the layer-space respectively.
We have chosen the Coulomb gauge $\nabla \cdot {\bf A}=0$ so that
the vertex corrections to $\hat{\bf \Gamma}$ are zero.
It is worth emphasizing that the vertex corrections in $\hat{\Gamma}_0$ are 
essential for preserving the gauge invariance of the theory. 

\begin{figure}
\begin{center}
\leavevmode
\epsfbox{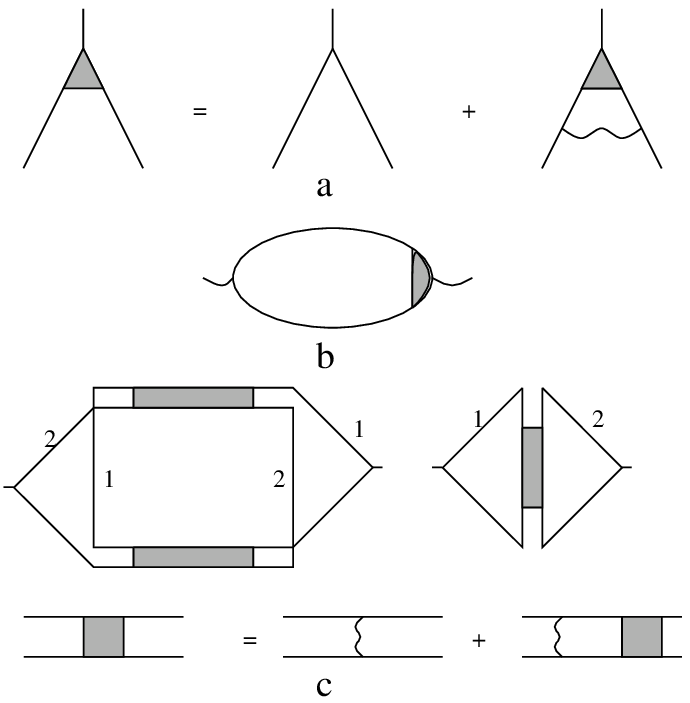}
\end{center}
\vspace{-0.4cm}
Fig.1 (a) Diagrams for
vertex corrections.
Solid lines represent the composite fermion 
Green functions $\hat{G}$ defined
in Nambu space; shaded triangles represent the renormalized
vertices. The wavy line stands for the irreducible 
interaction vertex.  
(b) Diagrams for the 
polarization of composite fermions.
(c) Diagrams for the drag conductivity  
above the critical temperature $T_c$.
Solid lines with index 1,2 are composite fermion
Green functions in layers 1,2 respectively.
\end{figure}
\vspace{-0.1cm}

The polarizability can be calculated in terms of the
diagrams in Fig.1b. Taking into account 
$\hat{G}, \hat{\Gamma},
\hat{{\bf \Gamma}}$ given in Eqs.2,4,
we obtain the results in the incompressible
paired quantum Hall liquid limit. 
At low temperatures $T \leq T_c$ in the limit $v_F Q \leq \Omega \ll \Delta$, 
we have 

\begin{eqnarray}
&&\pi^{dd}_{+}(i\Omega, {\bf Q}) 
={e^2} \frac{\partial n}{\partial \mu} 
[\alpha_1(T)\frac{{v_s}^2{Q}^2}{-\Omega^2 + {v_s}^2{Q}^2}
+\beta_2(T) \Xi_1(\Omega, {\bf Q})],
\nonumber \\
&&\pi^{dd}_{-}(i\Omega, {\bf Q}) 
={e^2} \frac{\partial n}{\partial \mu}\beta_1(T) 
\Xi_1(\Omega, {\bf Q})
\nonumber \\ 
&&\pi^{cc}_{+}(i\Omega, {\bf Q})=
- \frac{N_se^2}{m}[1-\beta_1(T)
\Xi_2(\Omega, {\bf Q})] 
\nonumber \\
&&\pi^{cc}_{-}(i\Omega, {\bf Q})=
- \frac{N_se^2}{m}[1-\alpha_1(T)-\beta_2(T) \Xi_2(\Omega, {\bf Q})] 
\end{eqnarray}
Here $\partial n/\partial\mu=m/2\pi$ is the thermodynamic
density of states, $m$ is the mass of composite 
fermions, and $N_s$ is the superfluid density. 
$\beta_1(T) \approx \beta_2(T)$
and $\beta_2(T)=1-\alpha_1(T)$; $\alpha_1(T)$ is given by 

\begin{eqnarray}
&&\alpha_1(T)=\left\{\begin{array}{cc}
1-2\sqrt{\frac{\pi T}{\Delta}}\exp(-\frac{\Delta}{T}),
\mbox{$T \ll T_c$}, \\
\frac{\pi\Delta}{4T}, \mbox{$T_c-T \ll T_c$}.
\end{array}\right. 
\end{eqnarray}
$\Xi_{1,2}$ have the following forms  
\begin{eqnarray}
\Xi_{1}=(\frac{\partial n}{\partial\mu})^{-1}
\sum_{\bf p} 
\frac{n_F(\epsilon_{{\bf p}+{\bf Q}}) -n_F(\epsilon_{\bf p})}
{-\Omega+\xi_{\bf p+Q}-\xi_{\bf p}+i\delta}
\nonumber \\
\Xi_{2}=\frac{2m}{N_s}\sum_{\bf p}v^2_F 
\frac{n_F(\epsilon_{{\bf p}+{\bf Q}}) -n_F(\epsilon_{\bf p})}
{-\Omega+\xi_{\bf p+Q}-\xi_{\bf p}+i\delta}
\end{eqnarray}
which were studied in detail in Ref.3.
When $\Omega \gg v_F Q$, $\Xi_{1,2}\propto Q^2/\Omega^2$.

The results in Eqs.5,6,7 can be interpreted in terms of
two-fluid model. $\pi^{dd}_+$ is the sum of the 
condensate contribution, which is proportional to
$\alpha_1$, and the quasiparticle contribution, which is
proportional to $\beta_{1,2}$. $\pi_+^{cc}$ is determined
mainly by the condensate component.
Asymmetrical polarizations $\pi^{cc}_-$, $\pi^{dd}_-$
have contributions mainly from thermally excited quasiparticles.
At $T\ll T_c$, the quasiparticle contributions are exponentially small
because of the energy gap in the spectrum. At temperatures close to $T_c$,
the condensate contribution becomes small.
Following Eqs.5,6,7, we find that $\sigma^{yy}_{11}=0$,
$\sigma^{xy}_{11}=\frac{e^2}{4 \pi \hbar}$ for this incompressible
paired quantum Hall state. The drag conductivity also vanishes in this
limit,  $\sigma^{yy}_{12}=\sigma^{xy}_{12}=0$ at $T < T_c$.
   
In the presence of random impurity potentials
$V_{1,2}({\bf r})$ in layer 1,2,
the composite fermions in different layers experience 
different random potentials. Composite
fermions in layer 1 have to pair with those in 
layer 2 with a different spectrum. 
In this case the Hamiltonian acquires additional terms;
${1 \over 2} (V_1({\bf r}) + V_2({\bf r})) 
(\psi^{\dagger}_1 \psi_1 + \psi^{\dagger}_2 \psi_2)
+ {1 \over 2} (V_1({\bf r}) - V_2({\bf r})) 
(\psi^{\dagger}_1 \psi_1 - \psi^{\dagger}_2 \psi_2)$. Here
$\psi_1, \psi_2$ are the composite fermion
operators in layers 1 and 2 respectively.
The second term acts
like a random Zeeman magnetic field 
on composite fermions
and effectively leads to the suppression of $\Delta$.
The impurity potentials pin
the Chern-Simons
flux in space\cite{HLR} and break the time reversal symmetry
of the composite fermion system.
This results in a further suppression $\Delta$.
Thus, in the strong disorder limit, the underlying composite
fermion system becomes gapless.

In the weak disorder
limit, $\tau\Delta \gg 1$, the energy gap of the quasiparticles
remains open. 
The change of the superfluid density $N_s$ and the sound velocity 
$v_s$ is proportional to $1/\tau\Delta$ and is negligible. 
However, the quasiparticle contributions are dramatically changed.
The longitudinal polarization $\Xi_1=DQ^2/i\Omega$ takes the 
diffusion form, 
while the transverse one $\Xi_2=\tau^{-1}/(i\Omega +\tau^{-1})$
is Drude-like.
Taking these into account, we find 

\begin{equation}
\sigma^{xx}_{12}=
\frac{\beta_{1}k_Fl}{1+\beta_1\beta_2(k_Fl)^2}\frac{e^2}{\hbar},
\sigma^{xy}_{12}=\frac{-1}{1+\beta_1\beta_2(k_Fl)^2}\frac{e^2}{8\pi\hbar}.
\end{equation}
in the weak disorder limit at $T \leq T_c$.
$\beta_{1,2}$ are given by Eq. 6, with $\Delta$ evaluated in
the presence of disorder. 
Since $\Delta$, which appears in the low temperature asymptotic
forms of $\beta_{1,2}$,
is a function of the elastic scattering rate $\tau^{-1}$ itself, the
drag conductivity as a function of temperature
strongly depends on disorder.
When $T$ becomes close to $T_c$, 
$\sigma^{xx}_{12}=(1/k_Fl)\times e^2/\hbar$
and $\sigma_{12}^{xy}=-(1/k_Fl)^2\times e^2/\hbar$.
On the other hand, they become exponentially small 
when $T$ goes to zero.
A similar temperature dependence of the drag conductivity was also
found in electron-hole double-layer system\cite{vignale}.
It is easy to confirm that, in both the pure and disordered
limit, the following equalities hold
\begin{equation}
\sigma^{xx}_{11}-\sigma^{xx}_{12}=0,
\sigma^{xy}_{11}-\sigma^{xy}_{12}=\frac{e^2}{4\pi\hbar}.
\end{equation}
Eq.9 can be attributed to the incompressibility 
of the paired quantum Hall state and does not
depend on disorder.

In the limit $\tau\Delta \ll 1$, the energy gap in the quasiparticle
spectrum disappears. 
In this case, $\beta_{1,2}$ become of order unity even at
$T=0$ and the exponential decay of the drag conductivity
at low temperatures does not occur.
As a result, for $T \leq T_c$,
$\sigma^{yy}_{12}\approx (1/k_Fl)\times e^2/\hbar$,
$\sigma^{xy}_{12} \approx (1/k_Fl)^2 \times e^2/\hbar$,
remaining finite even at zero temperature. 
It is worth pointing out that
in general $\Delta$ has energy dependence. 
However, note that $\pi^{dd}_+$ 
in Eq.5 manifests the existence of 
the Bogoliubov-Anderson mode in the
spontaneously symmetry-broken state and 
$\pi^{cc}_+$ reflects the
off-diagonal long range order in the composite fermion
system. Thus, Eq.8 follows as 
a consequence of the incompressibility of the paired quantum Hall 
state and does not depend on the detailed structure of $\Delta$. 

At high temperatures, the double-layer composite fermions are
weakly coupled with each other. However, when the 
critical temperature $T_c$ is approached,
the current-current polarizability 
diverges due to the strong pairing fluctuations of 
composite fermions in the two layers. This is similar to the 
situations discussed in\cite{Larkin}.
We find $\pi^{cc}_{12}(i\Omega, 0)
=i\Omega \sigma^{CF}_{12}$ in the $\Omega\rightarrow 0$ limit,
where

\begin{eqnarray}
&&\sigma^{CF}_{12} 
=\frac{e^2}{128 \hbar}\frac{\eta^2 D^2}{T^3} 
\int {d^2{\bf Q}}
\int^{+\infty}_0 d\Omega  
\frac{Q^2 Im  L^R( \Omega, {Q}^2)}
{\sinh^2\left(\frac{\Omega}{2T}\right)} 
\nonumber \\
&&[ Im L^R
+\frac{64 T^2\eta^{-2}D^{-2}} 
{\pi^3(l^{-2}+{Q}^{2})Q^2}
{Im\Psi(\frac{1}{2} +\frac{i\Omega +\eta D{Q}^2}{4\pi T})}].
\end{eqnarray}
Here $\eta=7\xi(3)/2\pi^3 T\tau$ and
$\Psi$ is the digamma function.  
The effective interlayer interaction 
is calculated in terms of the diagrams in Fig.1c,
\begin{eqnarray}
&&L^{R}(\Omega, {Q}^2)=\left(\frac{T-T_c}{T} +\frac{\pi}{8}
\frac{\eta D{Q}^2  + i\Omega}{T}\right)^{-1}.
\end{eqnarray}
We assume the temperature is close to $T_c$ and 
$T-T_c \ll \tau^{-1}$.
To leading order in $\tau(T-T_c)$, the contribution 
from the second term in Eq. 10 is negligible.
Taking into account Eqs. 1, 10 we obtain
\begin{eqnarray}
&&\sigma^{xx}_{12}=\frac{\pi}{4(k_F l)^2}\frac{T}{T-T_c}
\frac{e^2}{\hbar},
\ \
\sigma^{xy}_{12}=\frac{-\pi}{8(k_F l)^3}\frac{T}{T-T_c} \frac{e^2}{\hbar}.
\end{eqnarray}
Meanwhile $\sigma^{xx}_{11}=(2\pi/k_F l)\times e^2/\hbar$
and $\sigma^{xy}_{11}=e^2/4\hbar$ in zeroth order 
perturbation theory with respect to the pairing fluctuations.
When $T/(T -T_c) \sim k_F l$, $\sigma^{xx}_{12} \sim
\sigma^{xx}_{11}$ and the perturbation method breaks down.

In experiments, the drag resistivity is measured.
The drag resistivity tensor can be obtained by inverting the
conductivity tensor presented above.
At $T > T_c$, taking into account Eq.12, 
we get the corresponding drag resistivity  
\begin{eqnarray}
&&\rho^{xx}_{12}=\frac{\pi}{4(k_F l)^2}\frac{T}{T-T_c} \frac{\hbar}{e^2},
\end{eqnarray}
which {\em increases} as the temperature is decreased toward $T_c$.
The Hall drag resistivity is always zero in this model.
The contribution discussed in the 
previous papers\cite{kim,stern}, without
taking into account the contribution from the
pairing fluctuations,
is a monotonically {\em decreasing} function of temperature, i.e., 
$(l_BT/d\epsilon_F)^{4/3} \times \hbar/e^2$. 
Here $d$ is the interlayer spacing, assumed to be
larger than the magnetic length. 
Since the contribution due to pairing fluctuations 
{\em diverge} as $T_c$ is approached, we find that
as far as 
\begin{equation}
\frac{T}{T-T_c}\geq (k_Fl)^2 (\frac{l_BT}{d\epsilon_F})^{4/3}, 
\end{equation}
the result discussed here always overwhelms the 
contributions in\cite{kim,stern}.
$T_c$ is estimated as $(l_B/d)^2\epsilon_F$ in Ref.\cite{bonesteel}.
When $d \gg l_B$ and
$T_c \ll \epsilon_F$, Eq. 14 can be easily satisfied. 
Thus the drag resistivity
can develop a minimum as a function of temperature around $T_c$.

At $T < T_c$, following Eq.8, we obtain

\begin{equation}
\rho^{xx}_{11}=\rho^{xx}_{12}=\frac{2}{\beta_1(T)k_Fl}\frac{\hbar}{e^2} 
\end{equation}
which indicates that the drag resistivity diverges at low temperatures 
in the weak disorder limit when a gap still exists.
In the strong disorder limit, $\beta_1$ is of order unity even 
at $T=0$ and $\rho^{xx}_{12}$ remains finite at $T < T_c$.
We therefore suggest that the transition between the incompressible
paired quantum Hall state
and the weakly-coupled compressible
double-layer state could be responsible for 
the anomalous temperature dependence of the drag resistivity observed 
in the experiment\cite{eisenstein}.

In Ref.\cite{eisenstein}, 
no pronounced divergence was observed at low temperatures.
Instead,  
drag resistivity was shown to be saturated at low
temperatures.
This at first seems to indicate that a gapless limit was 
reached in the experiment.
In Ref.\cite{eisenstein}, 
$d\sim l_B \sim k_F^{-1} \sim 200{\rm \AA}$; the in-plane
longitudinal resistance is close to $3000$ Ohm and 
$l \sim k_F^{-1}$.
Indeed, this yields $\tau\Delta \sim 1$, implying a gapless
situation.
However,
to derive Eqs.8.9, we have assumed that, in the low temperature
phase, thermal fluctuations are negligible. 
When ${|T - T_c|}/{T_c} \leq {1}/{k_F l}$, 
fluctuations are strong and
the results in Eqs. 8, 13 are invalid. 
This sets the limit of the theory
when compared with the experiment {\em quantitatively}.
For the situation where $k_F l \sim 1$,
the transition regime where thermal
fluctuations are large could be of the same order as $T_c$.
It is plausible that the lowest temperature
in the experiment is still in the critical regime 
and the low temperature incompressible phase discussed 
in this paper was smeared out in
Ref.\cite{eisenstein}. 
To distinguish the gapless situation and
the thermal fluctuation effects, we suggest studying
double layer systems with $d \gg l_B$, where the
gapless limit can be reached ($\tau\Delta < 1$), while 
$\tau\epsilon_F$ is still greater than unity so that the
critical regime is narrow.   

Further complications arise
when the pairing wave function
also becomes inhomogeneous in space
in the presence of macroscopic inhomogeneities in 
the sample. 
The drag current is then carried by electron pairs
traveling along the percolating paths, which
are strongly dependent on impurity 
configurations and the amplitude of applied currents.
Finally, in the strong disorder limit, 
the mean field approach
is questionable due to strong quantum phase 
fluctuations present even at zero temperature.  
Solutions to these complications remain open.

Recently we became aware of a related work where 
the effect of the pairing fluctuation is 
also studied\cite{stern2}.
The discrepancy between some of the results in our 
initial manuscript and those of Ref.\cite{stern2} was
due to different boundary conditions\cite{stern3}.

We acknowledge the useful discussions
with I. Aleiner, B. Altshuler,
J. Eisenstein, H. Y. Kee, and especially
A. Stern. This work was supported by 
ARO under contract
DAAG 55-98-1-0270 (F.Z.) and 
NSF grant No. PHY9407194 (ITP at UCSB, Y.B.K.).

\end{multicols}


\begin{references}
\bibitem{eisenstein} M. P. Lilly, et.al,
Phys. Rev. Lett. {\bf 80}, 1714 (1998).
\bibitem{gramila} T. J. Gramila, et.al,
Phys. Rev. Lett. {\bf 66}, 1216 (1991).
\bibitem{macdonald} L.Zheng and A.H.MacDonald, Phys. Rev. B 
{\bf 48}, 8203 (1993).
\bibitem{kamenev} A. Kamenev and Y. Oreg, Phys. Rev. B {\bf 52},
7516 (1995).
\bibitem{HLR} B. I. Halperin, P. A. Lee, and N. Read, 
Phys. Rev. B {\bf 47}, 7312 (1993).
\bibitem{kim} Y. B. Kim, A. J. Millis, preprint, cond-mat/9611125.
\bibitem{stern}I. Ussishkin and A. Stern, Phys. Rev. B {\bf 56}, 4013 (1997).
\bibitem{sahki} S. Sakhi, Phys. Rev. B {\bf 56}, 4098 (1997).
\bibitem{Halperin} B. I. Halperin, Helv. Phys. Acta {\bf 56}, 75(1983).
\bibitem{Haldane} F. D. M. Haldane and E. H. Rezayi, Phys. Rev. Lett.
{\bf 60}, 956(1988).
\bibitem{Wen} M. Greiter, F. Wilczek, and X. -G. Wen, Phys. Rev. Lett.
{\bf 66}, 3205(1991).
\bibitem{bonesteel}N. E. Bonesteel, Phys. Rev. B {\bf 48}, 11484 (1993).
N. E. Bonesteel, I. A. MacDonald, C. Nayak,
Phys. Rev. Lett. {\bf 77}, 3009(1996). 
\bibitem{vignale}G. Vignale, A. H. MacDonald, Phys. Rev. Lett.
{\bf 76}, 2786 (1996).
\bibitem{AGD} A.Abrikosov, L.P.Gorkov, I.E.Dzyaloshinski,
{\it Methods of quantum field theory in statistical physics},
Dover, 1963.
\bibitem{Larkin}L. G. Aslamasov and A. I. Larkin, Fiz. Tverd. Tela.,
{\bf 10}, 1104 (1968)[Sov. Phys. Solid. State {\bf 10}, 975 (1968)];
K. Maki, Prog. Theor. Phys. {\bf 39}, 387 (1968);
R. S. Thompson, Phys. Rev. {\bf B 1}, 327 (1970).
\bibitem{stern2}I. Ussishkin, A. Stern, Phys. Rev. Lett.
{\bf 81}, 3932 (1998), reported also as cond-mat/9808013.
\bibitem{stern3}We are grateful to A. Stern for pointing this 
out to us.
\end{references}
\end{document}